\documentclass[mathleft]{an}
\usepackage{graphicx}
\usepackage{times}
\usepackage{natbib}
\bibpunct{(}{)}{;}{a}{}{,}

\newcommand{\oviii}{O\textsc{viii}}

\newcommand{\SWCX}{\textsc{swcx}}

\newcommand{\CXO}{\emph{Chandra}}
\newcommand{\CHIPS}{\textsc{chips}}
\newcommand{\EUVE}{\textsc{euve}}
\newcommand{\FUSE}{\textsc{fuse}}

\newcommand{\ROSAT}{\textsc{rosat}}

\newcommand{\swift}{\emph{Swift}}

\newcommand{\XMM}{XMM-\emph{Newton}}

\newcommand{\water}{\ensuremath{\textrm{H}_2\textrm{O}}}
\newcommand{\kms}{\ensuremath{\textrm{km~s}^{-1}}}

\newcommand{\Q}{mol\-e\-cules~s$^{-1}$}

\begin{document}

\Pagespan{789}{}
\Yearpublication{2012}%
\Yearsubmission{2012}%
\Month{02}%
\Volume{999}%
\Issue{88}%

\title{Cometary Charge Exchange Diagnostics in UV and X-ray}
\author{D. Bodewits\inst{1}\fnmsep\thanks{Corresponding author:
  \email{dennis@astro.umd.edu}\newline}
\and  D.J.~Christian\inst{2}
\and  J.A.~Carter\inst{3}
\and  K.~Dennerl\inst{4}
\and I.~Ewing\inst{2}
\and R.~Hoekstra\inst{5}
\and S.T~Lepri\inst{6}
\and  C.M.~Lisse\inst{7}
\and  S.J.~Wolk\inst{8}
}
\titlerunning{Charge Exchange Emission from Comets}
\authorrunning{D. Bodewits et al.}
\institute{
Dept. of Astronomy, University of Maryland, College Park, MD 20742-2421, USA 
\and 
Department of Physics  and Astronomy, California State University, 18111 Nordhoff Street, Northridge, CA 91330, USA
\and 
Department of Physics and Astronomy, University of Leicester, Leicester, LE1 1RH, UK
\and
Max-Planck-Institut fur extraterrestrische Physik,  Giessenbachstrasse, 85748 Garching, Germany
\and
KVI Atomic Physics, University of Groningen, Zernikelaan 25, NL-9747 AA Groningen, The Netherlands
\and
The University of Michigan, Ann Arbor, MI 48109Ð2143, USA
\and
Planetary Exploration Group,  Space Department, JHU-APL, 11100 Johns Hopkins Rd, Laurel, MD 20723, USA
\and
Harvard-Smithsonian Center for Astrophysics, 60 Garden  Street, Cambridge,  MA 02138, USA}

\received{14 Feb 2012}
\accepted{14 Feb 2012}
\publonline{later}


\abstract{%
Since the initial discovery of cometary charge exchange emission, more than 20 comets have been observed with a variety of X-ray and UV observatories. This observational sample offers a broad variety of comets, solar wind environments and observational conditions. It clearly demonstrates that solar wind charge exchange emission provides a wealth of diagnostics, which are visible as spatial, temporal, and spectral emission features. We review the possibilities and limitations of each of those in this contribution.}

\maketitle

\section{Introduction}

X-ray and Extreme Ultraviolet (EUV) emission is usually associated with high temperature environments. The discovery that comets are bright emitters in this spectral regime was therefore a big surprise \citep{Lis96,Mum97}, and is in strong contrast to our understanding that comets are dirty snowballs surrounded by a gaseous coma with a temperature of approximately 50~K. After the first discovery by \ROSAT{} of the X-ray emission from Comet C/1996 B2 (Hyakutake), a search through the observatory's archives proved that in fact all comets (with total visual magnitude m$_v < 12.0$) in the inner solar system ($\leq 2$~AU) had emitted X-rays \citep{Den97}. The total X-ray power in the 0.2 -- 1.0~keV band was between 0.004 -- 1.2~GW, the emission was highly variable in time, and many of the observed comets displayed a characteristic crescent shape. 

To explain these surprising observations, numerous possible scenarios were proposed. Amongst them were scattering/fluorescence of solar X-rays \citep{Kra97b}, thermal bremsstrahlung associated with collisions of solar wind electrons with cometary neutral gas or dust \citep{Bin97,Nor97,Uch98}, electron/proton K- and L-shell ionization \citep{Kra97b}, Rayleigh-scattering of solar X-rays by attogram dust particles \citep{Wic96,Owe98,Schu00}, and charge exchange between highly ionized solar wind minor ions and cometary neutral species \citep{Cra97}. A comparative study by \citet{Kra97b} demonstrated that none of these mechanisms except for the solar wind charge exchange emission and attogram dust scattering
 could account for the observed luminosities. The discovery of emission lines in the X-ray spectrum of Comet C/1999 S4 (\textsc{linear}) prove that the emission was driven by solar wind charge exchange \citep{Lis01}.

Before data from spacecraft monitoring the solar wind became available, observations of cometary ion tails were the only method of probing the solar wind. Even nowadays, comets largely remain the only means to sample regions outside the ecliptic plane. Cometary X-rays in particular have proven to be an excellent tool to study solar wind -- neutral gas interactions, because comets have no magnetic field and the wind therefore interacts directly with the neutral gas in the coma. Secondly, the size of the cometary atmosphere (on the order of 10$^5$~km) allows remote tracking of the ions as they penetrate into the comet's atmosphere, offering a close-up view on the interaction of the two plasmas. Thirdly, since the first observations of cometary X-ray emission, more than 20 comets have been observed with various X-ray and UV observatories including \ROSAT\ \citep{Lis96,Den97}, \EUVE\ \citep{Mum97}, BeppoSAX \citep{Owe98}, \CXO\ \citep{Lis01,Kra04b, Bod07, Wolk2009,Christian2010}, \textsc{chips}\ \citep{Sasseen2006}, \FUSE\ \citep{Feldman2005}, \XMM\ \citep{Schultz2006, Dennerl2012}, \emph{Suzaku}\ \citep{Brown2010}, and \swift\ \citep{Wil06,Carter2012}\footnote{Unsuccessful attempts to detect X-rays in comets were made using the Einstein satellite, the Rossi X-ray Timing Explorer, and the ASCA observatory \citep{Kra04b}.}. This observational sample offers a broad variety of comets, solar wind environments and observational conditions, clearly demonstrating that cometary charge exchange emission provides a wealth of diagnostics, which are visible as spatial, temporal, and spectral emission manifestations. We will discuss the possibilities and limitations of each of those in this contribution.

\section{Morphology}

\begin{figure}
\includegraphics[width=83mm]{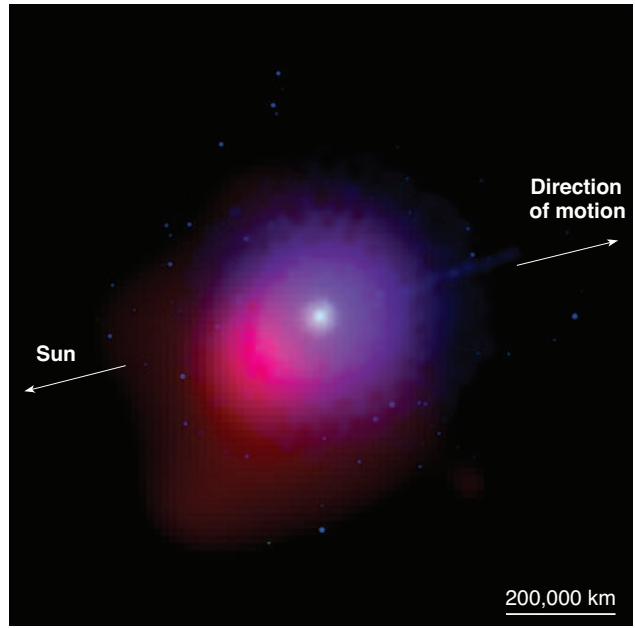}
\caption{\swift\ observations of comet 2007/N3 (Lulin), adapted from \cite{Carter2012}. The \swift\ satellite allows for simultaneous mapping of OH in the neutral coma (blue) and charge exchange X-ray emission (red). The peak X-ray brightness is offset by about 35,000 km towards the Sun.}
\label{label1}
\end{figure}

The emission morphology is a 3D 'tomography' of the solar wind flow through the neutral gas around the nucleus \citep{Weg04}. In very active comets (gas production rates of 10$^{29}$ \Q\ or more, \citealt{Bod07}) observed at phase angles of about 90 degrees, the X-ray emission maps a spherical gas distribution resulting in a characteristic sunward, crescent shape as seen by a remote observer, indicating a coma that is collisionally thick to charge exchange. A good example results from the combined observations of comet 2007/N3 (Lulin), shown in Figure~\ref{label1} \citep{Carter2012}. The \swift\  observatory is equipped with boresighted UV and X-ray instruments, allowing for simultaneous mapping of OH in the neutral coma (blue) and charge exchange X-ray emission (red). The peak X-ray brightness is offset by about 35,000 km towards the Sun. 

Less active comets (gas production rates of 10$^{28}$ \Q\ or less) have neutral comae that are virtually transparent to the incoming solar wind ions. The X-ray emission typically peaks at the nucleus, and jets or other local enhancements in the coma brighten, as was observed for the jet in 2P/Encke \citep{Lis05} and the curious X-ray morphology surrounding the fragmented comets C/1999 S4 (\textsc{linear}) and 73P/Schwass\-mann-Wach\-mann 3 \citep{Lis01,Wolk2009}. \cite{Weg05} analyzed a cometary x-ray image and deduced the shock location, but this has not yet been fully explored as their technique requires imaging data with a very high signal to noise ratio.

At 1 AU from the Sun, most of the charge exchange reactions take place in the denser parts of the coma within $\approx 10^5$ km from the nucleus \citep{Bod07}. For nearby comets, this often implies that the cometary emission extends well beyond the field of view and requires a careful background subtraction. This introduces an important problem for observers. The diffuse soft X-ray background is variable in time and place, and contains many of the \SWCX\ lines observed in comets resulting from charge exchange in the EarthÕs exosphere and the heliosheath \citep{Carter2008,Carter2011,Koutroumpa2007,Koutroumpa2008}. This background subtraction problem was explored in depth for the very extended Comet 17P/Holmes \citep{Christian2010}.

\section{Temporal Variation}

\begin{figure}
\includegraphics[width=83mm]{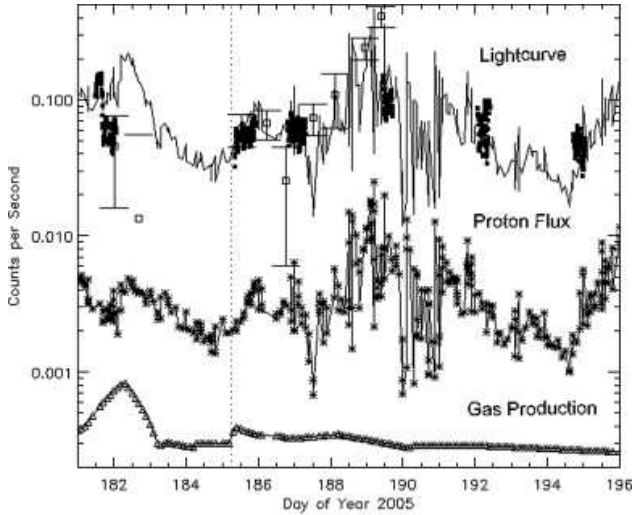}
\caption{The X-ray light curve is the result of comet activity and solar wind ion flux. Shown here are \CXO\ (filled squares) and \swift\ observations (open squares, \citealt{Wil06} -- normalized to the Chandra measurements) of Comet 9P/Tempel 1 around the time of Deep Impact. For comparison, we also show estimated water (gas) production rates (triangles, bottom), solar wind proton fluxes (stars, middle), and the product of these two, which is the predicted X-ray lightcurve (solid line, top). Figure adapted from \cite{Lis07}.}
\label{label2}
\end{figure}

Cometary X-ray luminosity is driven by a combination of the comet's gas production rate, the solar wind flux, and its heavy ion content. Furthermore, for nearby comets the effect of varying distance on the projected effective aperture should be taken into account. The interpretation of cometary X-ray light\-curves is therefore complex, as was first shown for Comet Hyakutake \citep{Neu00}. Long term \swift\ XRT and \CXO\ observations of Comet 9P/Tempel~1 \citep{Wil06,Lis07} demonstrated the richness of this approach  (see Fig.~\ref{label2}). Large increases in the X-ray luminosity could be assigned to a large cometary outburst (DOY 2005-180 to 184) followed by the interplanetary manifestation of a Coronal Mass Ejection (DOY 2005-188 to 190). A novel approach has recently been presented in \cite{Carter2012}, who directly constrained the cometary gas production rates and related this to the observed X-ray variability by employing \swift's co-aligned instrument suite, simultaneously observing comet C/2007 N3 (Lulin) in UV and X-rays. It is of note that the optical brightness of comets is driven by their dust content, which in itself is only weakly coupled to their overall gas activity \citep{AHearn1995,Jorda2008}. As the solar wind also has no visual manifestation in interplanetary space, there is therefore no simple relation between the optical and X-ray luminosities of comets.

The solar wind is highly variable in time and its ion composition can change dramatically over the course of less than a day due to variations in the solar source regions and dynamic changes in the solar wind itself. While the solar wind is sampled at the Earth's first Lagrangian Point L1 by several spacecraft, to accurately model \SWCX\ lightcurves solar wind properties have to be extrapolated to the position of the comet. Full MHD propagation models are available but are only applicable to comets in the equatorial plane. These models currently offer an accuracy of a day at best (see e.g. \cite{Fry03} and references therein), and only consider the wind's bulk properties, not its composition. A first order estimate of the state of the solar wind can be achieved using corotational mapping of the solar wind \citep{Neu00}, but this method only considers the Sun's rotation and the wind's bulk velocity, and performs poorly propagating dynamical structures in the wind. Latitudinal and corotational separations imply large inaccuracies in any solar wind mapping procedure and limit the comparison of solar wind data with X-ray observations to nearby comets \citep{Bod07,Wolk2009}. 

\section{Spectroscopy}
\begin{figure}
\includegraphics[width=75mm]{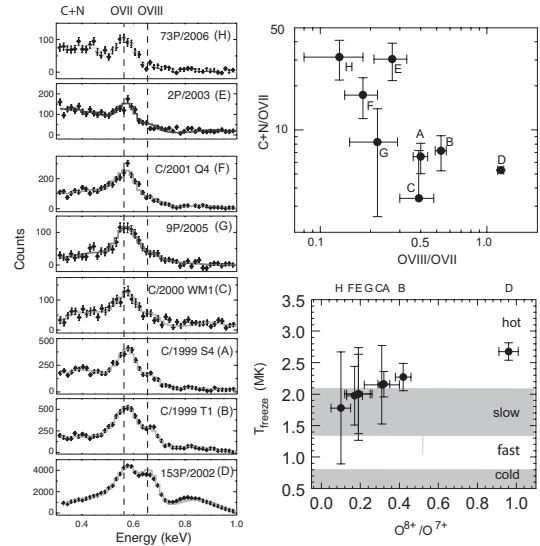}
\caption{Comparison of 8 different comets observed by \CXO\ between 2000 -- 2006, adapted from \citep{Bod07}. {\bf Left:} X-ray spectra, organized by color. {\bf Top right:} Flux ratios (colors) of the spectra. The low energy feature (C+N) is anti-correlated to the oxygen flux ratio. {\bf Bottom right:} Oxygen charge states derived from the \CXO{} spectra and their corresponding freeze-in temperatures. The shaded areas indicate typical freeze-in temperatures for different solar wind states.}
\label{label3}
\end{figure}

Charge exchange reactions are quasi-resonant processes, and depend strongly on properties of both the neutral and the ionized gas \cite{Cra02}. The resulting emission therefore provides a u\-nique window on their interactions. For example, within fusion experiments both Doppler shifts of charge exchange emission lines and their absolute and relative intensities are used to provide information on local plasma parameters such as temperatures, velocities and abundances and charge state of the interacting plasmas \citep{Isl94,Hel98,Hoe98,And00}. 

Current charge exchange investigations however are limited by observational constraints and sparse experimental data. Most cometary spectra have been obtained using \CXO 's \textsc{acis-s} instrument, which provides moderate energy resolution (110 eV \textsc{fwhm}) in the 300 to 2000 eV energy range. Several attempts have been made to extract ionic abundances from the X-ray spectra  \citep{Hae97,Weg98,Kha00,Sch00,Lis01,Kha01,Kra02,Weg04,Kra04a,Lis05}. However, the multitude of emission lines from hy\-dro\-gen- and helium like heavy ions in this regime implies that individual \SWCX{} lines are not resolved, often rendering least-square fits degenerate. More favorable approaches minimize the degrees of freedom by grouping all emission lines of each ion based on compilations of experimental and theoretical work \citep{Bei03,Bod07}.

\begin{figure}
\begin{center}
\includegraphics[width=75mm]{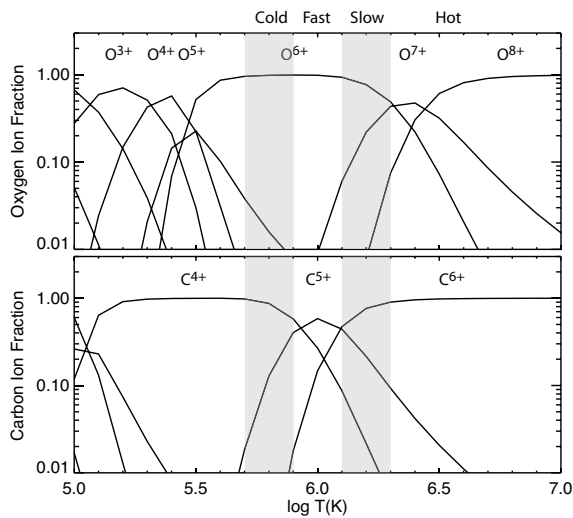}
\caption{Relative ion fractions for oxygen (top) and carbon (bottom), as a function of temperature in an optically thin plasma under thermal equilibrium, adapted from \cite{Bryans2009} and \cite{Landi2012}. Typical solar wind freeze-in temperatures are indicated by shaded areas.}
\label{label4}
\end{center}
\end{figure}

The overall shape of cometary \SWCX{} spectra is predominantly determined by the state of the solar wind \citep{Sch00,Kha01,Bod07}. During solar minimum the solar wind can be simplified to the stratification of a slow, low-latitude wind originating about the solar equator and a higher-latitude, faster wind originating from polar regions. In this bi-modal state the polar wind has a lower charge state than the wind at lower latitudes. This simple picture is complicated by the presence of transient phenomena (corotating interaction regions and coronal mass ejections), and by the chaotic state of the solar corona around solar maximum \citep{Gei95}. In low resolution spectra, these states manifest themselves as 'color differences', i.e. in the relative strength of the low energy ($<$500 eV), the \oviii\ emission at 561 eV and the \oviii\ emission at 653 eV\footnote{To avoid confusion, solar wind ions are in superscript notation and emitting ions are notated in linear notation.}. A comparison of 8 comets observed with \CXO{} showed a clear separation between the spectra that are dominated by the low energy component, and the 'oxygen-dominated' spectra. The spectra dominated by emission $>$500 eV show a gradual increase in the oxygen ionic ratio (Fig.~\ref{label3}). Because the solar wind is a collisionless plasma, this trend can be understood in terms of the freeze-in temperature of the solar wind source region (Fig.~\ref{label4}). The slow wind associated with streamers typically has freeze-in temperatures of 1.3 -- 2.1 MK \citep{Zur02,Bryans2009}. With lower freeze-in temperatures, such as found in solar wind associated with coronal holes or at higher solar latitudes, the oxygen charge state distribution decreases accordingly, and most oxygen is found in the \ensuremath{\textrm{O}^{6+}} state which emits in the Far UV after charge exchange. Because the low energy component of the spectrum contains both \ensuremath{\textrm{C}^{5+}} and \ensuremath{\textrm{C}^{6+}} which lines cannot be resolved with the current spectral resolution, it is less sensitive to changes in typical solar wind freeze-in temperatures. 


Due to its reliance on a good detection of the O~\textsc{viii} feature, the oxygen ion ratio has limited diagnostic use for fast and cool solar winds with freeze-in temperatures of 1~MK or less, such as the wind associated with coronal holes or the cold wind found at higher solar latitudes \citep{Christian2010}. Ion ratios between H- and He- like Carbon would make excellent probes for this temperature regime \citep{Landi2012} but their lines cannot be resolved at the spectral resolution currently achievable.  Extremely hot solar wind states provide another limiting case. The spectrum of Comet 153P/Ikeya-Zhang cannot be well explained by the \cite{Bod07} \SWCX{} model (Fig.~\ref{label3}). It anomalous spectrum was attributed to the interaction with the interplanetary manifestation of a coronal mass ejection (\textsc{icme}), likely resulting in many additional unresolved \SWCX{} lines of Fe~\textsc{xv--xx}. Both issues will be addressed once X-ray calorimeters become available to observers.

\subsection{High Energy - Above 1 KeV}
\SWCX{} X-ray spectra above 1 keV have not been well studied in comets, although tentative detections of Mg were presented in the spectrum of the exceptionally X-ray bright comet Ikeya-Zhang in \citet{Bod07}.  Also, \citet{Carter2009}  found evidence for Fe~\textsc{xvii -- xx}, Mg~\textsc{xi -- xii}, and Si~\textsc{xiii -- xiv} lines from the Earth's exosphere. Both papers attributed the presence of these highly charged ions due to coronal mass ejection events, which can have freeze-in temperatures of several millions of degrees \citep{Lep04}.

Recent work by \cite{Ewing2012} has confirmed the strong 1.3--2.0 keV emission lines from Ikeya-Zhang and detected lines at 1350, 1470, 1840 and 2000 eV that were identified as Mg~\textsc{xi}, Mg~\textsc{xii}, Si~\textsc{xiii}, and Si~\textsc{xiv} \SWCX{} lines. They demonstrated that upon close inspection of archival \CXO{} observations these Si and Mg lines were present at a significant level for at least 4 more of the comets.

\subsection{Low Energy - Below 300 eV} 
The close encounter with comet 73P/Schwass\-mann-\-Wach\-mann 3 (which came within 0.07 AU of Earth in 2006) allowed for (nearly) in situ measurements of the solar wind, enabling a direct test of spectral models. \cite{Wolk2009} compared ion abundances derived from \CXO{} observations with measurements by the \textsc{ace-swics} instrument. While the derived abundances of O$^{7+}$ and C$^{6+}$ were in good agreement with the solar wind data, the obtained relative abundance of C$^{5+}$ was off by an \emph{order of magnitude}. Part of this discrepancy might be explained by the lack of experimental charge exchange cross sections or calibration issues at the low-energy end of \CXO-\textsc{acis}. However, the excess low energy has been a persistent feature. \cite{Bod07} showed that the ratio between C$^{6+}$ and C$^{5+}$ ions and the oxygen ion ratio was poorly correlated, ostensibly because C$^{5+}$ abundances are consistently too large. It is therefore very likely that the source of the discrepancy is related to emission from several ions of species such as Fe, Mg, Si, Ne, and as much of 90\% of the emission around 300 eV should be attributed to species other than C$^{5+}$ \citep{Sasseen2006, Koutroumpa2008, Brown2010}.

\subsection{Ultraviolet}
\begin{figure}
\begin{center}
\includegraphics[width=70mm]{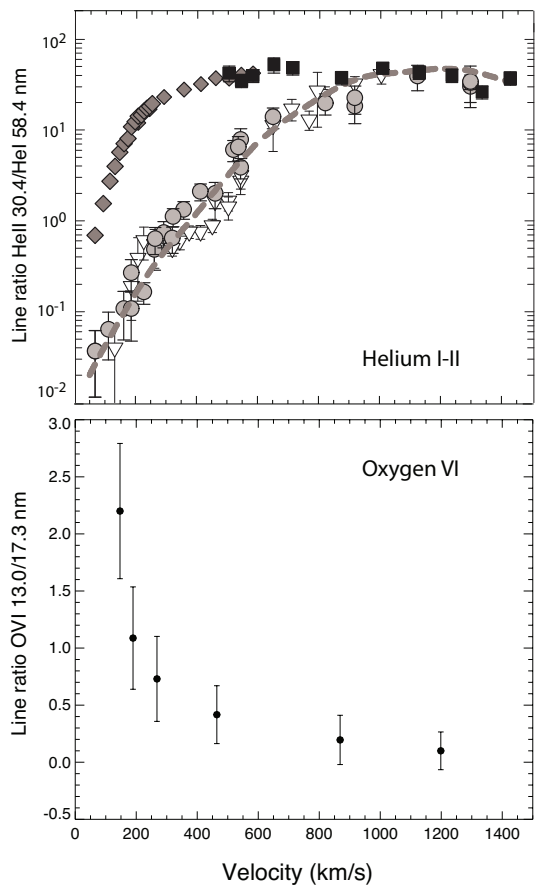}
\caption{Charge exchange diagnostics in the Extreme- and Far-UV. {\bf Top Panel:} Helium line ratios between 30.4 nm and 58.4 nm emission following He$^{2+}$ colliding on H$_2$ (white triangles), CO (grey circles), and \water{} (black squares and grey diamonds). {\bf Bottom Panel:} Velocity dependence of the O~\textsc{VI} line ratio between the 1s$^2$3d--1s$^2$2p and 1s$^2$4p--1s$^2$2s transitions at 17.3 and 13.0 nm, respectively. Figures adapted from \cite{Bod05,Bod07b}. }
\label{label5}
\end{center}
\end{figure}

As was noted above, many \SWCX{}\ lines fall below the bandwidth covered by current X-ray observatories. The excited states of H- and He-like C, N, and O decay in a radiative cascade which produces photons at X-ray, EUV, and UV wavelengths. In fact, the most abundant ions in the solar wind, He$^{2+}$ and O$^{6+}$, emit at EUV and FUV wavelengths. Model calculations based on O$^{q+}$ -- H$_2$ interactions underline the important role of O$^{6+}$ ions as line emitters \citep{Kha98,Liu99}. Both the \FUSE{} and \textsc{SOHO-UVCS} observatories were used to search for the O~\textsc{vi} doublet at 103 nm in five comets \citep{Wea02, Feldman2005,Raymond2002}. The non-detection of what are predicted to be the brightest \SWCX{} lines \cite{Kha01} might be attributed to the relatively small field of view of \FUSE{}, allowing observation of only a small fraction of the coma. As the four comets in these studies were all very active the O~\textsc{vi} brightness might have very well peaked outside \FUSE 's field of view \citep{Feldman2005, Sasseen2006}. The Cosmic Hot Interstellar Plasma Spectrometer (\CHIPS) spectrograph was used to observe three comets between 9 -- 26 nm, with a peak resolution of 0.4 nm \textsc{fwhm} \citep{Sasseen2006}. Interestingly, the authors reported a 2$\sigma$ detection O~\textsc{vi} in the spectrum of Comet C/2001 Q4 (\textsc{neat}). Experimental studies \citep{Bod07b,Miller2011} indicate that this  O~\textsc{vi} \SWCX{} emission between 10 -- 20~nm is another potent diagnostic of local solar wind velocities (Fig.~\ref{label5}). 

Helium ion are approximately 200 times more abundant than the oxygen ions that drive cometary X-rays. Low resolution \EUVE{} observations of several comets between 8 -- 70 nm allowed for the identification of O~\textsc{iii-vi}, C~\textsc{v}, and He~\textsc{i-ii} \SWCX{} emission lines \citep{Kra00,Kra01}. Helium charge exchange emission is dominated by two strong peaks; the He~\textsc{ii} (2p--1s) transition at 30.4 nm from one electron capture, and the He~\textsc{i} (1s2p--1s$^2$) transition at 58.4 nm following two-electron capture \citep{Bod04a,Bod05}. The ratio of these lines changes by more than three orders of magnitude for typical solar wind velocities between 100 to 1000~\kms (Fig.~\ref{label5}). This diagnostic was applied to \EUVE{} observations of comets Hale-Bopp and Hyakutake to interpret \SWCX{} emission in terms of local solar wind characteristics, providing an excellent window on what may be achieved once high resolution X-ray spectroscopy becomes available.

\section{Conclusion}
When solar wind ions pass through an atmosphere they are
neutralized via charge exchange reactions with the neutral gaseous
species. These reactions depend strongly on target species and
collision velocity. The resulting X-ray and UV emission is therefore a fingerprint of the underlying interaction, with many diagnostic qualities. 

To fully explore those diagnostics one has to consider all aspects relevant for cometary X-ray emission: experimental studies of state selective charge exchange cross sections and the resulting line emission, observational techniques to use X-ray telescopes that were certainly not designed to view objects within the solar system, and modeling of the propagation of the solar wind and its interaction with the gas in cometary comae. Together, these have greatly improved our understanding of the interaction of the solar wind with solar
system objects and in more general, of physical processes in
wind-environment collisions. The thorough understanding of
cometary charge exchange emission has opened the door to the
direct observation of more complex solar wind interactions such as
those with Earth, Mars, Venus, Jupiter, and the heliosphere.

\bibliographystyle{aa} 
\bibliography{references}

\begin{thebibliography}{65}
\expandafter\ifx\csname natexlab\endcsname\relax\def\natexlab#1{#1}\fi

\bibitem[{A'Hearn {et~al.}(1995)A'Hearn, Millis, Schleicher, Osip, \&
  Birch}]{AHearn1995}
A'Hearn, M.~F., Millis, R.~L., Schleicher, D.~G., Osip, D.~J., \& Birch, P.~V.
  1995, Icarus (ISSN 0019-1035), 118, 223

\bibitem[{{Anderson} {et~al.}(2000){Anderson}, {von Hellermann}, {Hoekstra},
  {Horton}, {Howman}, {Konig}, {Martin}, {Olson}, \& {Summers}}]{And00}
{Anderson}, H., {von Hellermann}, M., {Hoekstra}, R., {et~al.} 2000,
  Plasma~Phys.~Controll.~Fusion., 40, 781

\bibitem[{{Beiersdorfer} {et~al.}(2003){Beiersdorfer}, {Boyce}, {Brown},
  {Chen}, {Kahn}, {Kelley}, {May}, {Olson}, {Porter}, {Stahle}, \&
  {Tillotson}}]{Bei03}
{Beiersdorfer}, P., {Boyce}, K.~R., {Brown}, G.~V., {et~al.} 2003, Science,
  300, 1558

\bibitem[{{Bingham} {et~al.}(1997){Bingham}, {Dawson}, {Shapiro}, {Mendis}, \&
  {Kellet}}]{Bin97}
{Bingham}, R., {Dawson}, J.~M., {Shapiro}, V.~D., {Mendis}, D.~A., \& {Kellet},
  B.~J. 1997, Science, 275, 49

\bibitem[{Bodewits {et~al.}(2007)Bodewits, Christian, Torney, Dryer, Lisse,
  Dennerl, Zurbuchen, Wolk, Tielens, \& Hoekstra}]{Bod07}
Bodewits, D., Christian, D.~J., Torney, M., {et~al.} 2007, Astronomy and
  Astrophysics, 469, 1183

\bibitem[{Bodewits \& Hoekstra(2007)}]{Bod07b}
Bodewits, D. \& Hoekstra, R. 2007, Physical Review A, 76, 6

\bibitem[{{Bodewits} {et~al.}(2004){Bodewits}, {Juh{\'a}sz}, {Hoekstra}, \&
  {Tielens}}]{Bod04a}
{Bodewits}, D., {Juh{\'a}sz}, Z., {Hoekstra}, R., \& {Tielens}, A.~G.~G.~M.
  2004, ApJ.~Lett., 606, L81

\bibitem[{{Bodewits} {et~al.}(2005){Bodewits}, {Tielens}, {Morgenstern}, \&
  {Hoekstra}}]{Bod05}
{Bodewits}, D., {Tielens}, A.~G.~G.~M., {Morgenstern}, R., \& {Hoekstra}, R.
  2005, NIMB, 235, 358

\bibitem[{Brown {et~al.}(2010)Brown, Beiersdorfer, Bodewits, Porter, Ezoe,
  Hamaguchi, Hanya, Itoh, Kilbourne, \& Kohmura}]{Brown2010}
Brown, G., Beiersdorfer, P., Bodewits, D., {et~al.} 2010, The Energetic Cosmos:
  From Suzaku to Astro-H Otaru, Japan June 29, 2009 through July 2, 2009

\bibitem[{{Bryans} {et~al.}(2009){Bryans}, {Landi}, \& {Savin}}]{Bryans2009}
{Bryans}, P., {Landi}, E., \& {Savin}, D.~W. 2009, \apj, 691, 1540

\bibitem[{{Carter} {et~al.}(2012){Carter}, {Bodewits}, {Read}, \&
  {Immler}}]{Carter2012}
{Carter}, J., {Bodewits}, D., {Read}, A.~M., \& {Immler}, S.~M. 2012, A\&A, in
  press

\bibitem[{Carter {et~al.}(2010)Carter, Sembay, \& Read}]{Carter2009}
Carter, J., Sembay, S., \& Read, A. 2010, Monthly Notices of the Royal
  Astronomical Society, 402, 867

\bibitem[{Carter \& Sembay(2008)}]{Carter2008}
Carter, J.~A. \& Sembay, S. 2008, Astronomy and Astrophysics Supplement Series
  (ISSN 0365-0138), 489, 837

\bibitem[{Carter {et~al.}(2011)Carter, Sembay, \& Read}]{Carter2011}
Carter, J.~A., Sembay, S., \& Read, A.~M. 2011, Astronomy and Astrophysics,
  527, 115

\bibitem[{Christian {et~al.}(2010)Christian, Bodewits, Lisse, Dennerl, Wolk,
  Hsieh, Zurbuchen, \& Zhao}]{Christian2010}
Christian, D.~J., Bodewits, D., Lisse, C.~M., {et~al.} 2010, Astrophysical
  Journal Supplement Series, 187, 447

\bibitem[{{Cravens}(1997)}]{Cra97}
{Cravens}, T.~E. 1997, Geophys.~Res.~Lett., 24, 105

\bibitem[{{Cravens}(2002)}]{Cra02}
---. 2002, Science, 296, 1042

\bibitem[{{Dennerl} {et~al.}(1997){Dennerl}, {Englhauser}, \&
  {Tr\"{u}mper}}]{Den97}
{Dennerl}, K., {Englhauser}, J., \& {Tr\"{u}mper}, J. 1997, Science, 277, 1625

\bibitem[{{Dennerl}(2012)}]{Dennerl2012}
{Dennerl}, K. e.~a. 2012, in prep

\bibitem[{{Ewing} {et~al.}(2012){Ewing}, {Christian}, {Bodewits}, {Dennerl},
  {Lisse}, \& {Wolk}}]{Ewing2012}
{Ewing}, I., {Christian}, D., {Bodewits}, D., {et~al.} 2012, in prep.

\bibitem[{Feldman(2005)}]{Feldman2005}
Feldman, P.~D. 2005, Physica Scripta (ISSN 0031-8949), 119, 7

\bibitem[{{Fry} {et~al.}(2003){Fry}, {Dryer}, {Smith}, {Sun}, {Deehr}, \&
  {Akasofu}}]{Fry03}
{Fry}, C.~D., {Dryer}, M., {Smith}, Z., {et~al.} 2003, J.~Geophys.~Res., 108, 5

\bibitem[{{Geiss} {et~al.}(1995){Geiss}, {Gloeckler}, {von Steiger},
  {Balsiger}, {Fisk}, {Galvin}, {Ipavich}, {Livi}, {McKenzie}, {Ogilvie}, \&
  {Wilken}}]{Gei95}
{Geiss}, J., {Gloeckler}, G., {von Steiger}, R., {et~al.} 1995, Science, 268,
  1033

\bibitem[{{Haberli} {et~al.}(1997){Haberli}, {Gombosi}, {DeZeeuw}, {Combi}, \&
  {Powell}}]{Hae97}
{Haberli}, R.~M., {Gombosi}, T.~I., {DeZeeuw}, D.~L., {Combi}, M.~R., \&
  {Powell}, K.~G. 1997, Science, 276, 939

\bibitem[{{Hoekstra} {et~al.}(1998){Hoekstra}, {Anderson}, {Bliek}, {von
  Hellermann}, {Maggi}, {Olson}, \& {Summers}}]{Hoe98}
{Hoekstra}, R., {Anderson}, H., {Bliek}, F.~W., {et~al.} 1998,
  Plasma~Phys.~Controll.~Fusion., 40, 1541

\bibitem[{{Isler}(1994)}]{Isl94}
{Isler}, R.~C. 1994, Plasma Physics and Controlled Fusion, 36, 171

\bibitem[{Jorda {et~al.}(2008)Jorda, Crovisier, \& Green}]{Jorda2008}
Jorda, L., Crovisier, J., \& Green, D. W.~E. 2008, Asteroids, 1405, 8046

\bibitem[{{Kharchenko} \& {Dalgarno}(2000)}]{Kha00}
{Kharchenko}, V. \& {Dalgarno}, A. 2000, J.~Geophys.~Res., 105, 18351

\bibitem[{{Kharchenko} \& {Dalgarno}(2001)}]{Kha01}
---. 2001, ApJ.~Lett., 554, L99

\bibitem[{{Kharchenko} {et~al.}(1998){Kharchenko}, {Liu}, \&
  {Dalgarno}}]{Kha98}
{Kharchenko}, V., {Liu}, W., \& {Dalgarno}, A. 1998, J.~Geophys.~Res., 103,
  26687

\bibitem[{Koutroumpa {et~al.}(2007)Koutroumpa, Acero, Lallement, Ballet, \&
  Kharchenko}]{Koutroumpa2007}
Koutroumpa, D., Acero, F., Lallement, R., Ballet, J., \& Kharchenko, V. 2007,
  Astronomy and Astrophysics Supplement Series (ISSN 0365-0138), 475, 901

\bibitem[{Koutroumpa {et~al.}(2008)Koutroumpa, Lallement, Kharchenko, \&
  Dalgarno}]{Koutroumpa2008}
Koutroumpa, D., Lallement, R., Kharchenko, V., \& Dalgarno, A. 2008, arXiv,
  0805, 3212

\bibitem[{{Kras\-no\-pol\-sky}(1997)}]{Kra97b}
{Kras\-no\-pol\-sky}, V. 1997, Icarus, 128, 368

\bibitem[{{Krasnopolsky}(2004)}]{Kra04a}
{Krasnopolsky}, V.~A. 2004, Icarus, 167, 417

\bibitem[{{Krasnopolsky} {et~al.}(2002){Krasnopolsky}, {Christian},
  {Kharchenko}, {Dalgarno}, {Wolk}, {Lisse}, \& {Stern}}]{Kra02}
{Krasnopolsky}, V.~A., {Christian}, D.~J., {Kharchenko}, V., {et~al.} 2002,
  Icarus, 160, 437

\bibitem[{{Krasnopolsky} {et~al.}(2004){Krasnopolsky}, {Greenwood}, \&
  {Stancil}}]{Kra04b}
{Krasnopolsky}, V.~A., {Greenwood}, J.~B., \& {Stancil}, P.~C. 2004, Space
  Science Reviews, 113, 271

\bibitem[{{Krasnopolsky} \& {Mumma}(2001)}]{Kra01}
{Krasnopolsky}, V.~A. \& {Mumma}, M.~J. 2001, ApJ, 549, 629

\bibitem[{Krasnopolsky {et~al.}(2000)Krasnopolsky, Mumma, \& Abbott}]{Kra00}
Krasnopolsky, V.~A., Mumma, M.~J., \& Abbott, M.~J. 2000, Icarus (ISSN
  0019-1035), 146, 152

\bibitem[{{Landi} {et~al.}(2012){Landi}, {Alexander}, {Gruesbeck}, {Gilbert},
  {Lepri}, {Manchester}, \& {Zurbuchen}}]{Landi2012}
{Landi}, E., {Alexander}, R.~L., {Gruesbeck}, J.~R., {et~al.} 2012, \apj, 744,
  100

\bibitem[{{Lepri} \& {Zurbuchen}(2004)}]{Lep04}
{Lepri}, S.~T. \& {Zurbuchen}, T.~H. 2004, J.~Geophys.~Res., 109, 1112

\bibitem[{{Lisse} {et~al.}(2007){Lisse}, {Bodewits}, {Christian}, {Wolk},
  {Dennerl}, {Zurbuchen}, {Hansen}, {Hoekstra}, {Combi}, {Fry}, {Dryer},
  {M\"{a}kinen}, \& {Sun}}]{Lis07}
{Lisse}, C.~M., {Bodewits}, D., {Christian}, D.~J., {et~al.} 2007, Icarus, in
  press

\bibitem[{{Lisse} {et~al.}(2001){Lisse}, {Christian}, {Dennerl}, {Meech},
  {Petre}, {Weaver}, \& {Wolk}}]{Lis01}
{Lisse}, C.~M., {Christian}, D.~J., {Dennerl}, K., {et~al.} 2001, Science, 292,
  1343

\bibitem[{{Lisse} {et~al.}(2005){Lisse}, {Christian}, {Dennerl}, {Wolk},
  {Bodewits}, {Hoekstra}, {Combi}, {M{\"a}kinen}, {Dryer}, {Fry}, \&
  {Weaver}}]{Lis05}
---. 2005, ApJ, 635, 1329

\bibitem[{{Lisse} {et~al.}(1996){Lisse}, {Dennerl}, {Englhauser}, {Harden},
  {Marshall}, {Mumma}, {Petre}, {Pye}, {Ricketts}, {Schmitt}, {Trumper}, \&
  {West}}]{Lis96}
{Lisse}, C.~M., {Dennerl}, K., {Englhauser}, J., {et~al.} 1996, Science, 274,
  205

\bibitem[{{Liu} \& {Schultz}(1999)}]{Liu99}
{Liu}, W. \& {Schultz}, D.~R. 1999, Ap.J., 526, 538

\bibitem[{{Miller} {et~al.}(2011){Miller}, {Smith}, {Ehrenreich}, {Kessel},
  {Pollack}, {Verzani}, {Kharchenko}, {Chutjian}, {Lozano}, {Djuri{\'c}}, \&
  {Smith}}]{Miller2011}
{Miller}, K.~A., {Smith}, W.~W., {Ehrenreich}, T., {et~al.} 2011, The
  Astrophysical Journal, 742, 130

\bibitem[{{Mumma} {et~al.}(1997){Mumma}, {Krasnopolsky}, \& {Abbott}}]{Mum97}
{Mumma}, M.~J., {Krasnopolsky}, V.~A., \& {Abbott}, M.~J. 1997, ApJ.~Lett.,
  491, L125

\bibitem[{{Neugebauer} {et~al.}(2000){Neugebauer}, {Cravens}, {Lisse},
  {Ipavich}, {Christian}, {von Steiger}, {Bochsler}, {Shah}, \&
  {Armstrong}}]{Neu00}
{Neugebauer}, M., {Cravens}, T.~E., {Lisse}, C.~M., {et~al.} 2000,
  J.~Geophys.~Res., 105, 20949

\bibitem[{{Northrop} {et~al.}(1997){Northrop}, {Lisse}, {Mumma}, \&
  {Desch}}]{Nor97}
{Northrop}, T.~G., {Lisse}, C.~M., {Mumma}, M.~J., \& {Desch}, M.~D. 1997,
  Icarus, 127, 246

\bibitem[{{Owens} {et~al.}(1998){Owens}, {Parmar}, {Oosterbroek}, {Orr},
  {Antonelli}, {Fiore}, {Schultz}, {Tozzi}, {Maccarone}, \& {Piro}}]{Owe98}
{Owens}, A., {Parmar}, A.~N., {Oosterbroek}, T., {et~al.} 1998, ApJ.~Lett.,
  493, L47

\bibitem[{Raymond {et~al.}(2002)Raymond, Uzzo, Ko, Mancuso, Wu, Gardner, Kohl,
  Marsden, \& Smith}]{Raymond2002}
Raymond, J.~C., Uzzo, M., Ko, Y.-K., {et~al.} 2002, The Astrophysical Journal,
  564, 1054

\bibitem[{{Sasseen} {et~al.}(2006){Sasseen}, {Hurwitz}, {Lisse}, {Kharchenko},
  {Christian}, {Wolk}, {Sirk}, \& {Dalgarno}}]{Sasseen2006}
{Sasseen}, T.~P., {Hurwitz}, M., {Lisse}, C.~M., {et~al.} 2006, \apj, 650, 461

\bibitem[{Schultz {et~al.}(2006)Schultz, Owens, Rodr{\'\i}guez-Pascual, Lumb,
  Erd, \& St{\"u}we}]{Schultz2006}
Schultz, R., Owens, A., Rodr{\'\i}guez-Pascual, P.~M., {et~al.} 2006, Astronomy
  and Astrophysics Supplement Series (ISSN 0365-0138), 448, L53

\bibitem[{{Schulz} {et~al.}(2000){Schulz}, {St{\"u}we}, {Tozzi}, \&
  {Owens}}]{Schu00}
{Schulz}, R., {St{\"u}we}, J.~A., {Tozzi}, G.~P., \& {Owens}, A. 2000, A\&A,
  361, 359

\bibitem[{{Schwadron} \& {Cravens}(2000)}]{Sch00}
{Schwadron}, N.~A. \& {Cravens}, T.~E. 2000, ApJ, 544, 558

\bibitem[{{Uchida} {et~al.}(1998){Uchida}, {Morikawa}, {Kubotani}, \&
  {Mouri}}]{Uch98}
{Uchida}, M., {Morikawa}, M., {Kubotani}, H., \& {Mouri}, H. 1998, ApJ, 498,
  863

\bibitem[{{von Hellermann} {et~al.}(1991){von Hellermann}, {Mandl}, {Summers},
  {Boileau}, {Hoekstra}, {de Heer}, \& {Frieling}}]{Hel98}
{von Hellermann}, M., {Mandl}, W., {Summers}, H.~P., {et~al.} 1991,
  Plasma~Phys.~Controll.~Fusion., 33, 1805 1

\bibitem[{{Weaver} {et~al.}(2002){Weaver}, {Feldman}, {Combi}, {Krasnopolsky},
  {Lisse}, \& {Shemansky}}]{Wea02}
{Weaver}, H.~A., {Feldman}, P.~D., {Combi}, M.~R., {et~al.} 2002, ApJ.~Lett.,
  576, L95

\bibitem[{{Wegmann} \& {Dennerl}(2005)}]{Weg05}
{Wegmann}, R. \& {Dennerl}, K. 2005, A\&A, 430, L33

\bibitem[{{Wegmann} {et~al.}(2004){Wegmann}, {Dennerl}, \& {Lisse}}]{Weg04}
{Wegmann}, R., {Dennerl}, K., \& {Lisse}, C.~M. 2004, A\&A, 428, 647

\bibitem[{{Wegmann} {et~al.}(1998){Wegmann}, {Schmidt}, {Lisse}, {Dennerl}, \&
  {Englhauser}}]{Weg98}
{Wegmann}, R., {Schmidt}, H.~U., {Lisse}, C.~M., {Dennerl}, K., \&
  {Englhauser}, J. 1998, Planet.~Space~Sci., 46, 603

\bibitem[{{Wickramasinghe} \& {Hoyle}(1996)}]{Wic96}
{Wickramasinghe}, N.~C. \& {Hoyle}, F. 1996, Astroph.~Space~Sci., 239, 121

\bibitem[{{Willingale} {et~al.}(2006){Willingale}, {O'Brien}, {Cowley},
  {Jones}, {McComas}, {Mason}, {Osborne}, {Wells}, {Chester}, {Hunsberger},
  {Burrows}, {Gehrels}, {Nousek}, {Angelini}, {Cominsky}, {Snowden}, \&
  {Chincarini}}]{Wil06}
{Willingale}, R., {O'Brien}, P.~T., {Cowley}, S.~W.~H., {et~al.} 2006, ApJ,
  649, 541

\bibitem[{Wolk {et~al.}(2009)Wolk, Lisse, Bodewits, Christian, \&
  Dennerl}]{Wolk2009}
Wolk, S., Lisse, C.~M., Bodewits, D., Christian, D.~J., \& Dennerl, K. 2009,
  The Astrophysical Journal, 694, 1293

\bibitem[{{Zurbuchen} {et~al.}(2002){Zurbuchen}, {Fisk}, {Gloeckler}, \& {von
  Steiger}}]{Zur02}
{Zurbuchen}, T.~H., {Fisk}, L.~A., {Gloeckler}, G., \& {von Steiger}, R. 2002,
  Geophys.~Res.~Lett., 29, 66

\end{thebibliography}

\end{document}